\documentclass[english,aps,prl,twocolumn,amsmath,showpacs,superscriptaddress,preprintnumbers]{revtex4-1}
\usepackage{hyperref}

\usepackage{amsmath}
\usepackage{amssymb}
\usepackage{mathrsfs}
\usepackage{graphicx}
\usepackage{bm}
\usepackage{comment}

\def\cS{\mathcal{S}}

\newcommand\q{\mathbf{q}}
\newcommand\x{\mathbf{x}}

\newcommand\+{\dagger}
\newcommand\<{\langle}
\renewcommand\>{\rangle}
\renewcommand\d{\partial}

\begin{document}
	
	
\title{Algebraic Approach to Fractional Quantum Hall Effect}

\date{April 2018}
	
\author{Dung Xuan Nguyen}
\affiliation{Rudolf Peierls Center for Theoretical Physics, Oxford University, Oxford OX1 3PU, United Kingdom}
\author{Dam Thanh Son}
\affiliation{Kadanoff Center for Theoretical Physics, University of Chicago, Chicago, Illinois 60637, USA}
	
\begin{abstract}
We construct an algebraic description for the ground state and for the
static response of the quantum Hall plateaux with filling factor
$\nu=N/(2N+1)$ in the large $N$ limit.  By analyzing the algebra of
the fluctuations of the shape of the Fermi surface of the composite
fermions, we find the explicit form of the projected static structure
factor at large $N$ and fixed $z=(2N+1) q\ell_B\sim 1$. When $z<3.8$,
the result does not depend on the particular form of the Hamiltonian.
\end{abstract}
	
\maketitle    
	
\emph{Introduction.}---The most striking property of two-dimensional
electron gas in a strong magnetic field is its ability to form
nontrivial topologically ordered gapped states---the factional quantum
Hall (FQH) states~\cite{Tsui:1982yy}.  Despite the numerous
approaches, starting from the Laughlin wave
function~\cite{Laughlin:1983fy}, that have been brought to bear on the
problem, many important properties of the quantum Hall systems are
still beyond theoretical control of analytic approaches.  These
include the spectrum and dispersion of quasiparticle excitations, the
structure factors~\cite{Girvin:1986zz}, the electromagnetic and
gravitational response at finite wave number and frequency.
	
Generally, quantities that are not topologically protected should
depend on the details of the system under consideration, e.g., the
form of the electron-electron potential.  A few exceptions have
recently been found, suggesting that under some additional assumptions
(e.g., rotational invariance), some physical quantities are protected
by a combination of topology and symmetry.  For example, the $q^2$
correction to the Hall conductivity is related to the
shift~\cite{Hoyos:2011ez}.  It has also been argued that for a class
of chiral states and their particle-hole conjugates, the first two
terms in the small-$q$ expansion of the projected static structure
factor are uniquely defined by the filling factor, the shift, and the
central charge~\cite{Can:2014ota,Can:2014awa,Nguyen:2016hqh}.  On the
other hand, it seems that the higher terms (in $q$) of these
quantities are not universal, i.e., not determined solely by
topological characteristics of the quantum Hall state.
	
In this paper, we consider a regime in quantum Hall physics where the
full momentum dependence of the projected static structure factor,
i.e., \emph{all} terms in their momentum expansion, can be determined
in a reliable fashion.  The projected static structure factor is an
important property of the quantum Hall ground
state~\cite{Girvin:1986zz}.  We consider here quantum Hall plateaux at
filling factors $\nu=N/(2N+1)$ and their particle-hole conjugates
$\nu=(N+1)/(2N+1)$ (the Jain sequences).  The limit being considered
is that of large $N$, $N\gg1$, and finite $z=(2N+1)q\ell_B\sim1$,
where $q$ is the wavenumber at which the system is probed, and
$\ell_B$ is the magnetic length. The latter condition $z\sim1$
corresponds to wavelengths comparable to the cyclotron radius of the
composite fermion, which diverges when the filling factor approaches
$\nu=1/2$.  We will see that the momentum expansions of physical
quantities are in fact series expansions over $z$.  We will argue that
equal-time) structure factor~\cite{Girvin:1986zz} can be computed to
all orders in the expansion over $z$.  Our result for the projected
structure factor can be written as
\begin{equation}\label{result}
	\bar s(q) = \frac{N+1}8 (q\ell_B)^4 \, \frac {4 J_2(z)}{z J_1(z)},
\end{equation}
with
\begin{equation}
	z = (2N+1) q\ell_B .
\end{equation}
The result is valid as long as $z<z_1$, with $z_1\approx 3.83$ being
the first zero of the Bessel function $J_1$.  For $z>z_1$, $\bar s(q)$
cannot be predicted without knowing precise form of the Hamiltonian.
The divergence of the static structure factor at $z=z_1$ signals the
approach of the first magnetoroton minimum~\cite{Golkar:2016thq}, and
at finite $N$ this divergence should become a maximum of $\bar s(q)$.
	
We will argue that the result~(\ref{result}) is valid to the leading
order in $1/N$ in the scaling regime $qN\sim 1$.  We will also present
the indication that, for short-range electron-electron interaction,
the result is also valid to next-to-leading order in $1/N$, so the
correction to Eq.~(\ref{result}) is suppressed by $1/N^2$.
	
\emph{Algebra of shapes.}---Our understanding of the quantum Hall
effect near half filling is based on the concept of the composite
fermion~\cite{Jain:1989,Jain:1992,Zhang:1988wy,Fradkin:1991wy,Halperin:1992mh}.
In this paper, we use the modern, revised version of the composite
fermion field theory compatible with particle-hole
symmetry~\cite{Son:2015xqa}.  In this theory the composite fermion is
a massless Dirac fermion with a $\pi$ Berry phase around the composite
Fermi surface, and the emergent field has no Chern-Simons term in its
action.  It has been suggested that the emergence of the Dirac
composite fermion is a manifestation of a more general fermionic
particle-vortex duality~\cite{Wang:2015qmt,Metlitski:2015eka}.
	
Furthermore, following Refs.~\cite{Haldane:1994,Golkar:2016thq}, we
will interpret the low-energy excitations of the Dirac composite Fermi
liquid as fluctuations of the shape of the composite Fermi surface.
At each point in spacetime, the latter can be parametrized, in polar
coordinates in momentum space, as
\begin{equation}
	p_F(\theta) = p_F^0 + \sum_n e^{-in\theta} u_n .
\end{equation}
The scalar fields $u_n$ satisfy nontrivial commutation
relations \footnote{We use the complex notations $z=x+iy,
  \bar{z}=x-iy$},
\begin{multline}\label{un_comm}
	[u_n(\x), u_{n'}(\x')] = 2\pi\bigl(
	nb \delta_{n+n',0} \\ - i\delta_{n+n',1} \d_{\bar z}
	-i \delta_{n+n',-1} \d_z \bigr) \delta(\x-\x') .
\end{multline}
Here we have taken the external magnetic field $B=1$, so the magnetic
length $\ell_B=1$ \footnote{We chose natural units $c=\hbar=1$ and
  absorb the electron charge $e$ into $B$.}.  The average effective
magnetic field felt by the composite fermion is
$b=1/(2N+1)$ \footnote{$b=-1/(2N+1)$ corresponds to
  $\nu=(N+1)/(2N+1)$}.  The kinetic equation of Landau's Fermi liquid
theory at zero temperature can be reproduced by commuting $u_n$ with a
Lagrangian local in $u_n$, using the commutation
relations~(\ref{un_comm}).

In the case of the composite fermion of the fractional quantum Hall
states, the coupling of the composite fermion to a gauge field leads
to the freezing of $u_0$, $u_1$, and $u_{-1}$, corresponding to the
density and currents of the composite fermions.  This can be seen most
clearly if one assumes that the gauge field $a_\mu$ has no kinetic
term.  This case extremization of the action with respect to $a_\mu$
the vanishing of the composite fermion density fluctuations and
current, and so of $u_0$ and $u_{\pm1}$.  We will argue later on that
the presence of a kinetic term for $a_\mu$ only affects our
calculations to subleading orders in $1/N$.
	
Eliminating $u_0$ and $u_{\pm1}$, we see that in Eq.~(\ref{un_comm}),
the only nonzero commutators are between $u_n$ and $u_{n'}$ with
indices $n$ and $n'$ having opposite signs.  We can rewrite this
equation as
\begin{equation}
	[u_n(\q),\, u_{-n'}(-\q')] = C_{nn'}(\q) \delta_{\q\q'}
	\quad n,n' \ge 2 ,
\end{equation}
where $C_{nn'}$ is a tridiagonal matrix given by
\begin{equation}
	C_{nn'}(\q) = 2\pi \left( nb \delta_{n,n'} 
	+ q_{\bar z} \delta_{n,n'+1} 
	+ q_z \delta_{n,n'-1} \right) ,
\end{equation}
or
\begin{equation}
	C = 2\pi \begin{pmatrix} 2b & q_z & 0 & 0 & \ldots \\
	q_{\bar z} & 3b & q_z & 0 & \ldots \\
	0 & q_{\bar z} & 4b & q_z & \ldots \\
	0 & 0 & q_{\bar z} & 5b & \ldots \\
	\ldots & \ldots & \ldots & \ldots & \ldots
	\end{pmatrix} .
\end{equation}

	
\emph{Canonical pairs.}---At $q\neq0$, the commutation relation
between $u_n$ does not have the canonical form.  To bring it into the
canonical form, one performs a unitary transformation
\begin{equation}
	v_n(\q) = \sum_{m=2}^\infty U_{nm}(\q) u_m(\q) ,
	\end{equation}
so that
\begin{equation}
	[v_n(\q),\, v_{-n'}(-\q')] = \tilde C_{nn'}(\q)
	\delta_{\q\q'} ,
\end{equation}
where $\tilde C_{nn'}$ is now a diagonal matrix with coefficient
	\begin{equation}
	\left\{
	\begin{array}{cc} \frac{2\pi}{2N+1}\lambda_n, & n=n', \\ 0, & n\neq n' .
	\end{array}
	\right.
	\end{equation}
Here $\lambda_n$ \footnote{$\lambda_n$ is the eigenvalue that
  corresponds to the eigenvector $v_n$.} are the solutions to the
characteristic equation
\begin{equation}\label{recursion}
	(m-\lambda)b x_m + q_{\bar z} x_{m-1} + q_z x_{m+1} = 0, \quad m\ge2 .
	\end{equation}
Without losing generality we can take $q_x=q$, $q_y=0$.  The solution
to the recursion relation (\ref{recursion}), for $m\ge3$ is
\begin{equation}
	x_m = (-1)^m [\alpha J_{m-\lambda}(z) + \beta Y_{m-\lambda}(z)], \quad z=\frac qb \,.
\end{equation}
In order for $x_m$ to decrease at $m\to\infty$ we have to set
$\beta=0$.  Then Eq.~(\ref{recursion}) is satisfied for $m=2$ only
when
\begin{equation}\label{eigenval}
	J_{1-\lambda}(z) =0 , \qquad z = (2N+1)q.
\end{equation}
This equation determines the eigenvalues of the matrix $C$.  When $z$
is smaller than the first zero of the Bessel function $J_1$,
$z_1\approx3.8$, all solutions to Eq.~(\ref{eigenval}) (as an equation
for $\lambda$) are positive.  In this case, one can define the
creation and annihilation operators $a_n$ and $a^\+_n$ as follows
	\begin{equation}
	a_n = \sqrt{\frac{2N+1}{2\pi\lambda_n }} {v_n}\,, \quad
	a_n^\+ =  \sqrt{\frac{2N+1}{2\pi\lambda_n }} {v_{-n}}\,.
	\end{equation}
The commutator of $a_n$ and $a_n^\+$ now has the canonical form,
	\begin{equation}
	[ a_n(\q),\, a^\+_{n'}(\q')] =
	\delta_{nn'}  \delta_{\q\q'} .
	\end{equation}
In contrast, when $z>z_1$, some of $\lambda_n$ are negative.  For
these eigenvalues, $v_{-n}$ should be interpreted as an annihilation
operator ($v_{-n}\sim a_n$), while $v_n$ is a creation operator
($v_n\sim a_n^\+$).
	
\emph{Hamiltonian.}---In order to find the ground state one needs a
Hamiltonian.  We will argue later it is sufficient to consider a
Hamiltonian quadratic in $u_n$,
	\begin{equation}\label{Hamiltonian}
	H =
	\sum_{\q}
	\sum_{m,n=-\infty}^\infty
	h_{mn}(\q) u_{-m}(-\q) u_n(\q).
	\end{equation}
Due to rotational invariance $h_{mn}\sim q^{|m-n|}$.  Since
$q\sim1/N$.  This means that, to leading and next-to-leading orders in
$1/N$, the only nonzero elements of $h_{mn}$ are those with $m-n=0$ or
$\pm1$.  In particular $h_{mn}=0$ when $m$ and $n$ are of opposite
sign: the first term in the Hamiltonian of this type is $h_{-2,2}u_2
u_2$, where $h_{-2,2}\sim q^4$ and is suppressed by a high power of
$1/N$.  We thus will assume that $h_{mn}$ are nonzero only when $m$
and $n$ are of the same sign.  This turns out to be the only
information about the Hamiltonian that we will need.
	
One can transform $H$ to be a quadratic form of $a_n$,
\begin{equation}
	H =
	\sum_{\q}
	\sum_{m,n=2}^\infty H_{mn} a_m^\+(\q) a_n(\q).
\end{equation}
The spectrum of excitations then can be obtained by diagonalizing the
matrix $H_{mn}$.  The ground state is the state that is annihilated by
the $a_n$,
	\begin{equation}
	a_n |0\> = 0.
	\end{equation}
	For $(2N+1)q<z_1$, this means $v_n|0\>=0$, which also means that 
	\begin{equation}
	u_n |0\> = 0.
	\end{equation}
	
\emph{Charge density.}---Now we continue with our discussion of the
density operator.  In the composite fermion field theory, it is
identified with the magnetic field of the emergent gauge field:
$\delta\rho=-\delta b/(4\pi)$ \cite{HLR,Son:2015xqa,Nguyen:2017dcf}.
We need to relate $b$ to $u_n$ to make further progress.  To do that,
first we note that the equation of motion for $u_1$ is
\cite{Nguyen:2017qck,Nguyen:2017dcf}
	\begin{equation}
	\dot u_1 = e_{\bar z} + i[u_1,\, H] = e_{\bar z} - 2\pi\d_z
	\frac{\delta H}{\delta u_2} \,, 
	\end{equation}
	with $e_i=\partial_i a_0-\partial_0 a_i$. Physically the term $e_{\bar z}$ comes from the acceleration of the
	composite fermion moving in an emergent electric field.  Setting
	$u_1=0$ we obtain
	\begin{equation}\label{ebarz}
	e_{\bar z} = 2\pi  \d_z \frac{\delta H}{\delta u_2} \,.
	\end{equation}
	Analogously
	\begin{equation}\label{ez}
	e_z = 2\pi \d_{\bar z} \frac{\delta H}{\delta u_{-2}} \,.
	\end{equation}
	
We will combine Eqs.~(\ref{ebarz}) and (\ref{ez}) with the Bianchi
identity,
	\begin{equation}\label{Bianchi}
	\delta \dot b + 2 (\d_z e_{\bar z} - \d_{\bar z} e_z ) = 0,
	\end{equation}
to find $\delta b$.  This seems to still require the knowledge of the
Hamiltonian.  We now show that this is not true, and we can express
$b$ in terms of $u_n$ without knowing the explicit form of the
Hamiltonian.  We now note that the equation of motion for $u_n$, with
$n\ge2$, is
\begin{equation}
\dot u_n = -i[u_n,\, H] = -i [u_n,\, u_{-m}] \frac{\delta H}{\delta u_{-m}} ,
\end{equation}
In this and subsequent equations, the indices $n$, $m$ runs from 2 to
$\infty$ and summation over repeating indices is implied.  In momentum
space,
\begin{equation}
	u_n(\q) = -i C_{nm} (\q) \frac{\delta H}{\delta u_{-m}(-\q)} .
\end{equation}
Inverting this equation, we can write
\begin{equation}
	\frac{\delta H}{\delta u_{-n}(-\q)} = iC^{-1}_{nm}(\q) \dot u_m(\q) ,
\end{equation}
where $C^{-1}_{nm}$ are the elements of the matrix inverse of $C$.
Analogously we find
\begin{equation}
	\frac{\delta H}{\delta u_n(-\q)} = -i u_{-m}(\q) C^{-1}_{mn}(-\q) .
\end{equation}
This allows us to write $e_{\bar z}$ and $e_z$ as
\begin{align}
	e_{\bar z} (\q) &= -2\pi q_z C^{-1}_{2m}(\q) \dot u_m(\q),\\
	e_z(\q) &= 2\pi q_{\bar z} \dot u_{-m}(\q) C^{-1}_{m2}(-\q) .
\end{align}
Substituting this equation into the Bianchi identity (\ref{Bianchi})
and integrating over time, we then find the density of electrons in
terms of the shape fluctuations
\begin{equation}\label{rho-u2}
	\delta\rho(\q) = -q_z^2 C^{-1}_{2m}(\q) u_m(\q) - q_{\bar{z}}^2 u_{-m}(\q) C^{-1}_{m2}(-\q) .
\end{equation}
As an example, consider the regime of very small $q$, $z\ll1$.  In
this regime the only nonzero component of $C^{-1}$ is
$C^{-1}_{22}=(4\pi b)^{-1}$, and Eq.~(\ref{rho-u2}) becomes
\begin{equation}
	\delta\rho = \frac{2N+1}{4\pi}(\d_z^2 u_2 + \d_{\bar z}^2 u_{-2}) .
\end{equation}
This matches with the result found in in the bimetric theory of the
nematic phase transition~\cite{Gromov:2017qeb,Nguyen:2017qck}, where
$u_{\pm2}$ are identified with two independent components of a
unimodular dynamical metric~\cite{Haldane:2009ke,Haldane:2011ia} and
$\delta\rho$ is proportional to the Gaussian curvature constructed
form this metric.
	
We can now use the equation just derived to find the equal-time
correlation of the density, $\<\delta\rho \delta\rho\>$.  Note that if
$q$ is such that $(2N+1)q$ is smaller than the first zero of the
Bessel function $J_1$, the ground state is annihilated by $u_n$ with
$n>0$: $u_n|0\>=\<0|u_{-n}=0$.  From that we find
\begin{multline}
	\< \delta\rho (\q) \delta\rho (-\q)\> = \frac{q^4}{16}
	C^{-1}_{2m}(\q) [u_m(\q),\, u_{-l}(-\q)] C^{-1}_{l2}(\q)\\
	= \frac{q^4}{16} C^{-1}_{22}(\q) .
\end{multline}
To find $C^{-1}_{22}$, one needs to solve the equation $C_{nm}
x_m=\delta_{n2}$ and read out $x_2$.  Using the same technique which
we used to find the eigenvalues of $C$, we find
\begin{equation}
	C_{22}^{-1}(\q) = \frac1{4\pi b} \frac{4 J_2(z)}{z J_1(z)} \,.
\end{equation}
	
The result is then
\begin{equation}
	\< \delta\rho (\q) \delta\rho (-\q)\> = \frac{(2N+1)q^4}{64\pi}
	\frac{4 J_2(z)}{z J_1(z)} \,.
\end{equation}
Finally, after dividing by the density of electrons, we find the
projected structure factor (neglecting sub-sub-leading terms in $1/N$)
\begin{equation}\label{sq-result}
	\bar s(q) = \frac{(N+1)q^4}8 \, \frac{4J_2(z)}{z J_1(z)} \,.
\end{equation}
Note that this form of $\bar s(q)$ has been found previously near a
nematic phase transition~\cite{Nguyen:2017qck}; here we have
demonstrated that this formula is valid in general.
	
\emph{Discussion of the result.}---We now expand $s(q)$ at small $q$,
\begin{equation}\label{sq-expanded}
	\bar s(q) = \frac{N+1}8 q^4 + \frac{N^2(N+2)}{48} q^6 + \cdots
\end{equation}
	The $q^4$ coefficient of the static structure factor $\bar
	s(q)=s_4q^4+ s_6 q^6$ saturates the Haldane
	bound~\cite{Haldane:2009ke}
	\begin{equation}
	s_4\equiv \lim_{q\rightarrow 0}\frac{\bar{s}(q)}{(q\ell_B)^4} =
	\frac{|\cS-1|}{8},
	\end{equation}
where $\cS$ is the shift~\cite{WenZee,Read:2008rn}, which is equal to
$N+2$ for the state under consideration.  The saturation of Haldane
bound was derived earlier for a large class of trial
states~\cite{Nguyen:2014uea,Read:2007cv}.  However, Jain's sequences
do not belong to those states, so the saturation of the Haldane bound
is a nontrivial fact.  Note, however, that Jain's states share with
the states considered in Ref.~\cite{Nguyen:2014uea,Read:2007cv} the
property of chirality: all edge modes propagate in one direction.
	
We now consider $s_6$.  Little is know about the constraints on $s_6$.
In contrast to $s_4$, there is no constraint analogous to the Haldane
bound for $s_6$.  Nevertheless, in
Refs.~\cite{Can:2014ota,Can:2014awa,Nguyen:2016hqh} it has been argued
that for certain chiral states (including the Laughlin and Pfaffian
states), $s_6$ is completely determined by topological characteristics
of the quantum Hall state.  Extending the result of
Ref.~\cite{Can:2014ota,Can:2014awa,Nguyen:2016hqh} to states where
the composite fermions fill multiple Landau levels, one
finds~\cite{Gromov:2017qeb}
\begin{equation}
	\bar s_6 = \frac1{8\nu} \left( \nu\textrm{var}(s)
	+ \nu\varsigma^2 - \nu\varsigma + \frac{\nu-c}{12}\right),
\end{equation}
	where $\varsigma$ is the guiding center orbital spin of the composite
	fermion and $\textrm{var}(s)$ is the orbital spin variance \cite{Gromov:2015vars}.  For the
	Jain's state with $\nu=N/(2N+1)$, only the first two terms contribute
	to leading and subleading orders in $N$:
	$\nu\textrm{var}(s)=\frac1{12}N(N^2-1)$ and $\varsigma=\frac12(N+1)$,
	so $\bar s_6=\frac1{48}N^3(N+2)$, in agreement with
	Eq.~(\ref{sq-expanded}).
	
Equation~(\ref{sq-result}) can also be obtained from the dynamic
structure factor in a direct but not very illuminating manner.  We
have performed a (rather cumbersome) calculation of the dynamic
structure factor by using the equations of motions for $u_n$ in a
theory with $H$ having the form (\ref{Hamiltonian}) with the diagonal
elements given by the Landau parameters in the presence of a space-
and time-varying external scalar potential $A_0$.  We confirm that the
dynamic structure factor has a complicated structure with an infinite
number of poles corresponding to an infinite number of neutral
excitations.  After integrating over frequency, we
find~\cite{NS-unpublished} that, rather miraculously, the infinite sum
over residues can be taken exactly for $z<z_1$, the Landau parameters
drop out, and the result matches with
Eq.~(\ref{sq-result})~\footnote{The result has previously been quoted
  in Ref.~\cite{Nguyen:2016hqh}.}.
	
\emph{Sources of corrections.}---In our treatment we have set $u_0$
and $u_{\pm1}$ to zero, which means that we have assumed the absence
of the kinetic term for $a_\mu$ in the Lagrangian.  One source of
corrections to the result~(\ref{sq-result}) is the kinetic term of
$a_\mu$.  In the Dirac composite fermion theory the Chern-Simons term
for $a_\mu$ is forbidden by particle-hole symmetry, and for
short-ranged electron-electron interaction the kinetic term appears
first at order $(\d_i a_j)^2\sim N^{-2} a^2$ when $q\sim1/N$.  In
contrast, the loop induces a term $\frac\omega q a^2$.  The regime we
are interested in is $q\sim 1/N$ and $\omega\sim1/N^z$, where $z$ is
the dynamic critical exponent of a fermion coupled to a U(1) gauge
field (up to three loops $z=3/2$~\cite{Metlitski:2010pd}, but the
result may be changed in higher loops~\cite{Holder:2015hya}), so the
fermion-loop induced term is $\sim N^{1-z}a^2$.  The bare term is
suppressed by $N^{3-z}$ relative to the fermion loop
contribution. Assuming $z<2$ the kinetic term for $a_\mu$ can be
neglected without affecting the result to leading and next-to-leading
orders in $1/N$.
	
We now try to estimate corrections due to interactions between the
modes $u_n$.  A treatment of this technically difficult problem is
deferred to future work.  We can, however, make the following
observation.  The action for $u_n$ can be expected to have the form
\begin{equation}
	\mathcal L \sim N u\d_t u + \frac1{N^{z-1}} (u^2 + u^3 + u^4 +
        \cdots).
\end{equation}
Here the $N$ dependence of the coefficient of the $u\d_t u$ is
dictated by the commutation relations (\ref{un_comm}), that of the
$u^2$ term is determined by the energy gap, which is expected to be
$O(N^{-z})$, and the coefficients of the interactions terms ($u^3$,
$u^4$) come from the assumption that nonlinear effects in energy
appear when $u\sim1$.  We need estimate the magnitude of the loop
diagrams.  There are two one-loop contributions to the $u$
self-energy.
	
	\begin{figure}[t]
		\includegraphics[width = 0.4\columnwidth]{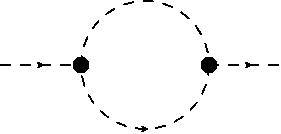}
		\hspace{0.1\columnwidth}
		\includegraphics[width = 0.27\columnwidth]{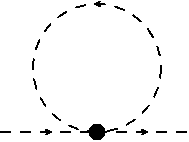} 
		\caption{One-loop contribution to $u$ self-energy.}
		\label{fig:1loop}
	\end{figure}
	
First, internal lines with energy much larger than the gap ($N^{-z}$)
simply lead to renormalization of the low-energy effective theory for
$u_n$.  Thus, it is sufficient to limit oneself to $O(1)$ number of
internal modes with the energy of order the gap.  Second, we can
restrict ourselves to internal momentum where non-Fermi liquid effects
are expected to be important.  At energy scale $\omega$ this occurs
for $q\lesssim\omega^{1/z}$, in our case, $q\sim N^{-1/2}$.  The first
diagram in Fig.~\ref{fig:1loop} then is of order
\begin{equation}
	\left( \frac1{N^{z-1}}\right)^2\! \int\! d\omega\, dq \left( N^{z-1}\right)^2  \sim \frac1{N^{z-1}} \frac1{N} \,. 
\end{equation}
This is by a factor of $1/N^2$ smaller than the inverse bare
propagator for $u$.  Analogously, the second diagram in
Fig.~\ref{fig:1loop} can be estimated in the same way to be suppressed
by $1/N^2$.
	
When the interaction of the electron is not short-ranged, but Coulomb,
the the situation is different.  We can estimate the influence of the
Coulomb interaction by comparing the bare Coulomb interaction $|q|
a_i^2$ with the one-loop contribution to the gauge self-energy, which
is of order $\frac\omega q a^2\sim a^2$ (ignoring powers of $\ln N$).
The bare action is of order $1/N$ compared the term generated by the
fermion loop.  This means the results will be modified to order $1/N$.
The result~(\ref{sq-result}) remains valid at leading order in $1/N$
but one expects corrections to first order in $1/N$.
	
\emph{Conclusions}.---In this paper we have shown that the projected
structure factor can be computed reliably for states on Jain's
sequences near $\nu=1/2$, and the result depends on the kinematic
structure of the Fermi surface of the composite fermion rather than
detailed knowledge of the Hamiltonian.  As the result, even when
composite fermions at $\nu=1/2$ form a non-Fermi liquid, 
we can still
compute $\bar s(q)$ reliably, although in a regime of momentum which
shrinks to zero as the filling factor approaches $1/2$.  It would be
interesting to see if the algebraic method used here is useful in a
wider range of problems related to the non-Fermi liquid.

The result~(\ref{sq-result}) is amenable to numerical verification.
One should keep in mind that in the more directly measurable
\emph{unprojected} static structure factor $s(q)=1-e^{-q^2/2}+\bar
s(q)$, the contribution from the projected structure factor $\bar
s(q)$ is subdominant in $1/N$.  Recent numerical
data~\cite{Balram:2017yth} are consistent with Eq.~(\ref{sq-result})
at small $z$, but also (and even a bit better) with a truncation of
$\bar s(q)$ to the sum of the $q^4$ and $q^6$ terms.  It may be that
larger $N$ is needed to clearly see the presence of a $q^8$ terms in
$\bar s(q)$.

We thank Andrey Gromov and Max Metlitski for discussions.  D.T.S.  is
supported, in part, by DOE grant No.\ DE-FG02-13ER41958 and a Simons
Investigator grant from the Simons Foundation.  D.X.N. is supported,
in part, by EPSRC grant EP/N01930X/1.
	
\bibliography{SSF}
	
\end{document}